%% file: main.tex
\documentclass{INTERSPEECH2023}

\interspeechcameraready

\title{Multi-microphone Automatic Speech Segmentation in Meetings Based on Circular Harmonics Features}
\name{Théo Mariotte$^{1,2}$, Anthony Larcher$^2$, Silvio Montrésor$^1$, Jean-Hugh Thomas$^1$}
\address{
  $^1$LAUM UMR CNRS 6613 IAGS, Le Mans Université, France\\
  $^2$LIUM, Le Mans Université, France}
\email{theo.mariotte@univ-lemans.fr, anthony.larcher@univ-lemans.fr}

\usepackage{comment}
\usepackage{adjustbox}
\begin{document}

\maketitle
 
\begin{abstract}

Speaker diarization is the task of answering \textit{Who spoke and when?} in an audio stream. 
Pipeline systems rely on speech segmentation to extract speakers' segments and achieve robust speaker diarization. 
This paper proposes a common framework to solve three segmentation tasks in the distant speech scenario: Voice Activity Detection (VAD), Overlapped Speech Detection (OSD), and Speaker Change Detection (SCD). 
In the literature, a few studies investigate the multi-microphone distant speech scenario. 
In this work, we propose a new set of spatial features based on direction-of-arrival estimations in the circular harmonic domain (CH-DOA).
These spatial features are extracted from multi-microphone audio data and combined with standard acoustic features.
Experiments on the AMI meeting corpus show that CH-DOA can improve the segmentation while being robust in case of deactivated microphones.

\end{abstract}
\vspace{-10pt}
\input{01_intro_v2.tex}

\input{011_tasks.tex}

\input{02_spat_feat.tex}
\input{03_protocol.tex}
\input{04_seg_results.tex}

\vspace{-5pt}
\input{06_ccl.tex}

\newpage
\bibliographystyle{IEEEtran}
\bibliography{mybib}

\end{document}

%% file: 01_intro_v2.tex
\section{Introduction}

Speaker diarization is the task of answering \textit{Who spoke and when?} in an audio stream~\cite{anguera2012speaker,park2022review}.
Many speaker diarization approaches are based on pipeline architectures~\cite{park2022review,bredin_pyannoteaudio_2020,landini2022bayesian}.
Those approaches rely on a speech segmentation step that extracts speaker-homogeneous segments. 
Speaker clustering is then performed by extracting and grouping speaker embeddings via clustering algorithms~\cite{park2022review}.
This paper focuses on automatic speech segmentation, which can be divided into three sub-tasks: Voice Activity Detection (VAD), Overlapped Speech Detection (OSD), and Speaker Change Detection (SCD).

VAD detects speech segments in the audio signal. 
It is the first step in most speaker diarization pipelines~\cite{park2022review}.
Finally, since overlapping speech is one of the major sources of errors in speaker diarization pipelines~\cite{bredin21_interspeech,garcia2019speaker}, OSD is required.
It consists in detecting speech segments in which multiple speakers are simultaneously active.
SCD is also required to detect speaker turns in the audio signal, i.e., when the currently active speaker is changing.

Early studies on VAD~\cite{sohn1999statistical,ng2012developing}, OSD~\cite{boakye_improved_2011,charlet_impact_2013,yella2014overlapping} and SCD~\cite{siegler1997automatic,liu1999fast} are based on the statistical modeling of acoustic features.
The latter is originally solved by comparing the statistics of two adjacent segments. 

Statistical models have then been replaced by neural networks due to their strong modeling capacities.
VAD~\cite{bredin21_interspeech,ryant2013speech,lavechin2019end}, OSD~\cite{bredin21_interspeech,bullock_overlap-aware_2020,Cornell2022}, and SCD~\cite{bredin2017tristounet,yin2017speaker,hruz2017convolutional} can be solved by modeling a sequence of acoustic features and performing a frame-level binary classification.
SCD is also tackled as the regression of functions in which maxima are located at turn locations~\cite{hruz2017convolutional}.




Most VAD, OSD, and SCD studies are conducted on single-channel data. 
In the meeting context, recording signals with a distant device offers practical benefits since it does not require participants to carry an individual microphone.
Microphone arrays are commonly used as distant devices to capture additional spatial information.
Few studies have been conducted on multi-microphone speech segmentation~\cite{Cornell2022,hu2015speaker,cornell_detecting_2020,mariotte22_interspeech}.
In particular, Cornell~\textit{et al.}~\cite{Cornell2022} explore the use of Interaural Phase Difference (IPD) spatial features for joint VAD and OSD and report a noticeable performance gain.
Hu~\textit{et al.}~\cite{hu2015speaker} investigate the use of Time Difference of Arrival (TDOA) features to detect speaker changes. 
Although the authors show diarization performance gain, experiments were only conducted on simulated data.
To the best of our knowledge, no other work has been reported on the use of spatial features for distant SCD. 

Several spatial features have been investigated in various multi-microphone speech processing tasks~\cite{pak2019sound,vincent2018audio,gu2020enhancing,grumiaux2021high}.
In particular, SongGong~\textit{et al.}~\cite{songgong_robust_2021} propose a speaker localization method based on circular harmonics (CH).
Although these features require the use of a circular array, they depend little on the number of available microphones.
Hence, circular harmonics are an interesting framework for feature extraction to rely less on the array configuration.

In this paper, we tackle VAD, OSD, and SCD tasks with the same architecture.    
We propose the use of CH to extract spatial features for multi-microphone speech segmentation.
This choice is motivated by the common use of circular microphone arrays to capture distant speech in meetings~\cite{carletta2006ami,fu21b_interspeech}.
The proposed spatial features consist of direction-of-arrival estimation in the CH framework (CH-DOA).
Spatial features are combined with commonly used acoustic features to solve segmentation tasks.
As far as authors are aware, this is the first investigation on the use of CH features for distant speech segmentation.
Furthermore, we report the impact of IPD spatial features for SCD since no work has been found considering their use for this task.
We demonstrate that adding spatial information drastically improves the detection of speaker turns.
Finally, we present encouraging results of CH-DOA-based OSD and SCD systems under mismatched array conditions.
The code will be available soon in a large-scale diarization toolkit\footnote{\url{https://git-lium.univ-lemans.fr/speaker}}.

The paper is organized as follows. Sect.~\ref{sect:tasksssssssssssssss} presents VAD, OSD, and SCD tasks. Sect.~\ref{sect2_ch_feat} presents the CH-DOA feature extraction.
Sect.~\ref{sect3_protocol} introduces the speech segmentation model along with the dataset and the experimental protocol before presenting results in Sect.~\ref{sect4_seg_perf} and conclusions in Section~\ref{sect_ccl}.

%% file: 011_tasks.tex
\section{Segmentation Tasks}

\label{sect:tasksssssssssssssss}

\begin{figure}[ht!]
    \centering
    \includegraphics[width=0.9\linewidth]{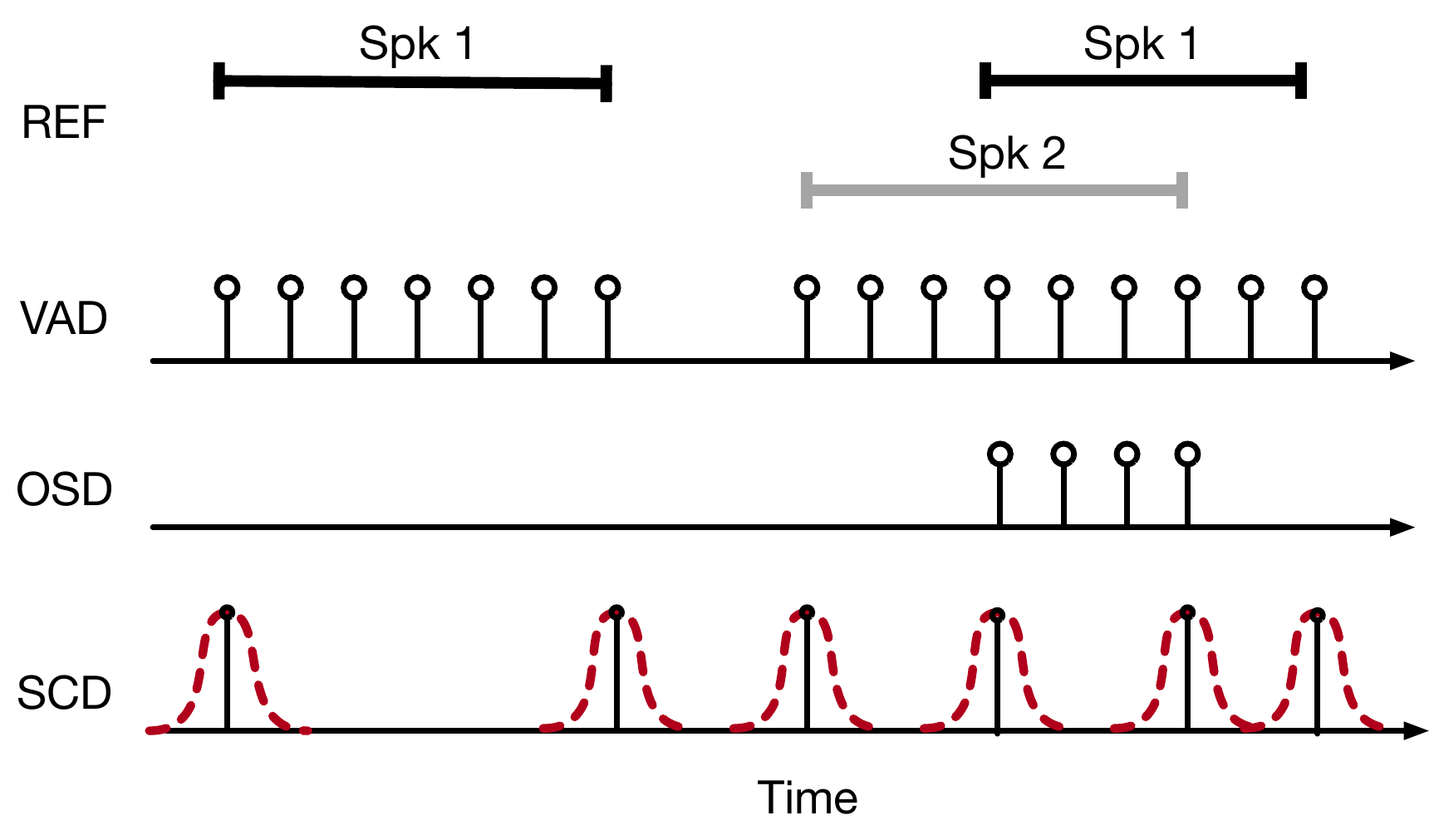}
    \caption{VAD, OSD, and SCD labels along with the reference diarization. VAD and OSD are formulated as a binary frame classification task while SCD is solved as the regression of Gaussian-like functions.}
    \label{fig:labels}
\end{figure}

This section describes the segmentation tasks considered. 
The labeling procedure for each task is presented in figure~\ref{fig:labels}.

\vspace{-15pt}

\subsection{Voice Activity and Overlapped Speech Detection}

VAD is formulated as a framewise binary sequence classification task~\cite{bredin_pyannoteaudio_2020}.
Let $\boldsymbol{X}=[\boldsymbol{X}_0,\dots,\boldsymbol{X}_t,\dots,\boldsymbol{X}_{T-1}]$ be a sequence of features extracted from the audio signal with $T$ being the length of the sequence.
The VAD task aims at predicting the sequence $\boldsymbol{P}=[\boldsymbol{P}_0,\dots,\boldsymbol{P}_t,\dots,\boldsymbol{P}_{T-1}]$, where $\boldsymbol{P}_t = \{p(N_{spk}=0|\boldsymbol{X}_t),p(N_{spk}\geq1|\boldsymbol{X}_t)\}$ is the pseudo-probability of the $t$-th frame to belong to each class.
$N_{spk}$ denotes the number of active speakers in the current frame.


OSD formulation is similar to VAD.
However, positive labels correspond to frames containing strictly more than one active speaker: $\boldsymbol{P}_t = \{p(N_{spk}\leq1|\boldsymbol{X}_t),p(N_{spk}\geq2|\boldsymbol{X}_t)\}$.

\subsection{Speaker Change Detection}

In this paper, SCD is formulated as a regression task~\cite{hruz2017convolutional}.
This approach consists in estimating functions in which maxima are located at speaker change points.
Speaker changes are represented as Gaussian-like functions\footnote{The variance $\sigma^2$ of the Gaussian functions used as labels is randomized during training. It follows a uniform distribution $\mathcal{U}$ such as $\sigma^2 \sim \mathcal{U}_{[2,7]}$.} following the binary label encoding of~\cite{he_deep_2018}. 
In this work, the speech-to-non-speech transition is considered as a speaker turn (Fig.~\ref{fig:labels}).

%% file: 02_spat_feat.tex
\section{Circular Harmonics Features}
\label{sect2_ch_feat}

Circular harmonics (CH) are a set of elementary 2-d functions.
They are similar to cylindrical harmonics or spherical harmonics in the 3-d domain~\cite{williams2000fourier}.
An acoustic signal can be represented as a weighted sum of CH components.
This section presents this formulation and introduces the CH-DOA features used for speech segmentation.

\subsection{Circular harmonics framework}

Let us consider a uniform circular microphone array (UCA) composed of $M$ microphones with a radius $r$.
The $X_m(f,t)$ signal captured by the $m$-th microphone can be expressed in the short-time Fourier transform (STFT) domain as a weighted sum of circular harmonics~\cite{torres_room_2016}: 

\begin{equation}
    X_m(f,t) = \sum_{n=-\infty}^\infty C_n(f,t) e^{jn\phi},
    \label{eq:pressure_as_ch}
\end{equation}
where $f$ denotes a frequency bin and $t$ the frame index.
In equation~\eqref{eq:pressure_as_ch}, $j=\sqrt{-1}$, $\phi$ is the DOA of the sound source, $e^{jn\phi}$ is the $n$-th order CH and $C_n(f,t)$  the associated coefficient.
By using a circular microphone array, the sound field is sampled at some discrete locations.
CH coefficients are estimated as follows~\cite{songgong_robust_2021,torres_room_2016}:

\begin{equation}
    \tilde{C}_n(f,t) = \frac{1}{M}\sum_{m=1}^M X_m(f,t)e^{-jn\psi_m},
\end{equation}
where $\tilde{C}_n(f,t)$ is the estimated CH coefficient and $\psi_m = (m-1)\frac{2\pi}{M}$ denotes the angle of the $m$-th microphone.

\subsection{CH-DOA feature extraction}

Spatial filtering, i.e. beamforming, can be performed in the CH domain~\cite{torres_room_2016}.
This is also known as \textit{modal beamforming} and can be expressed as follows~\cite{songgong_robust_2021}:

\begin{equation}
    B_n(f,t) = \sum_{-N}^N\frac{\tilde{C}_n(f,t)}{j^nJ_n(kr)}e^{jn\theta},
    \label{eq:beamforming_modal}
\end{equation}
with $B_n(k,t)$ being the $n$-th order beamformed signal and $k=2\pi f/c$ the wave number with $c$ the speed of sound. 
$J_n(kr)$ is the $n$-th order Bessel function of the first kind and $\theta$ indicates the steering direction.

The Pseudo-Intensity Vector (PIV) uses only zero- and first-order beamformers. 
The zero-order beam is obtained from equation~\eqref{eq:beamforming_modal} by setting $N=0$:

\begin{equation}
    B_0(f,t) = \frac{\tilde{C}_0(f,t)}{J_0(kr)}.
\end{equation}

For $N=1$, two orthogonal beams can be defined oriented towards the $\theta_x=0$ and $\theta_y=\pi/2$ respectively. 
The beam $B_{1x}(f,t)$ (respectively $B_{1y}$ with $\theta_y$) is expressed following:
\begin{equation}
    B_{1x}(f,t) = \sum_{-1}^1\frac{\tilde{C}_n(f,t)}{j^nJ_n(kr)}e^{jn\theta_{x}}.
\end{equation}

Then, the PIV components $I_x$ and $I_y$ can be calculated as:

\begin{equation}
    \bigg [\begin{aligned}
    I_{x}(f,t)\\
    I_{y}(f,t)
    \end{aligned}
\bigg ] = \frac{1}{2}\mathfrak{R}\Bigg\{B_0^{*}(f,t)
    \bigg [
    \begin{aligned}
    B_{1x}(f,t)\\
    B_{1y}(f,t)
    \end{aligned}
    \bigg ]
    \Bigg\}
\end{equation}
where $\mathfrak{R}$ denotes the real part and $^*$ the complex conjugate.

The PIV is supposed to be oriented in the propagation direction of the impinging acoustic wave.
Thus, the angular direction of the PIV in the frame of reference of the microphone corresponds to the source DOA~\cite{songgong_robust_2021}:
\begin{equation}
    \hat{\phi}(f,t) = \arctan\Bigg\{\frac{I_y(f,t)}{I_x(f,t)}\Bigg\}.
    \label{eq:doa_est}
\end{equation}

The estimated DOA $\hat{\phi}$ is used as a spatial feature for speech segmentation and is denoted as CH-DOA.
The following section presents how these features are integrated into our segmentation systems.
The CH-DOA features offer a similar computational complexity as IPD/CSIPD since it only relies on the multi-microphone STFT without any additional loop.





%% file: 03_protocol.tex
\section{Experimental protocol}
\label{sect3_protocol}

This section presents the experimental protocol to evaluate the impact of CH-DOA spatial features on multi-microphone speech segmentation.

\subsection{Dataset}

Experiments are conducted on the AMI meeting corpus~\cite{carletta2006ami}.
This dataset is about 100h of speech acquired during realistic meetings. 
The majority of participants are non-native English speakers and were asked to conduct a design project. 
Speech can be either spontaneous or scripted depending on the session.
Meetings have been recorded using various devices. 
Experiments are conducted on the AMI \textit{Array 1} data, which is an 8-microphone circular array placed in the center of the table.
Training, development, and evaluation partitions follow the protocol proposed in~\cite{landini2022bayesian}.
Labels for VAD, OSD, and SCD are extracted from the manual annotation of the segments.
Speech signals are sampled at 16kHz.

\subsection{Segmentation architecture}
\begin{figure}[ht]
    \centering
    \includegraphics[width=0.9\linewidth]{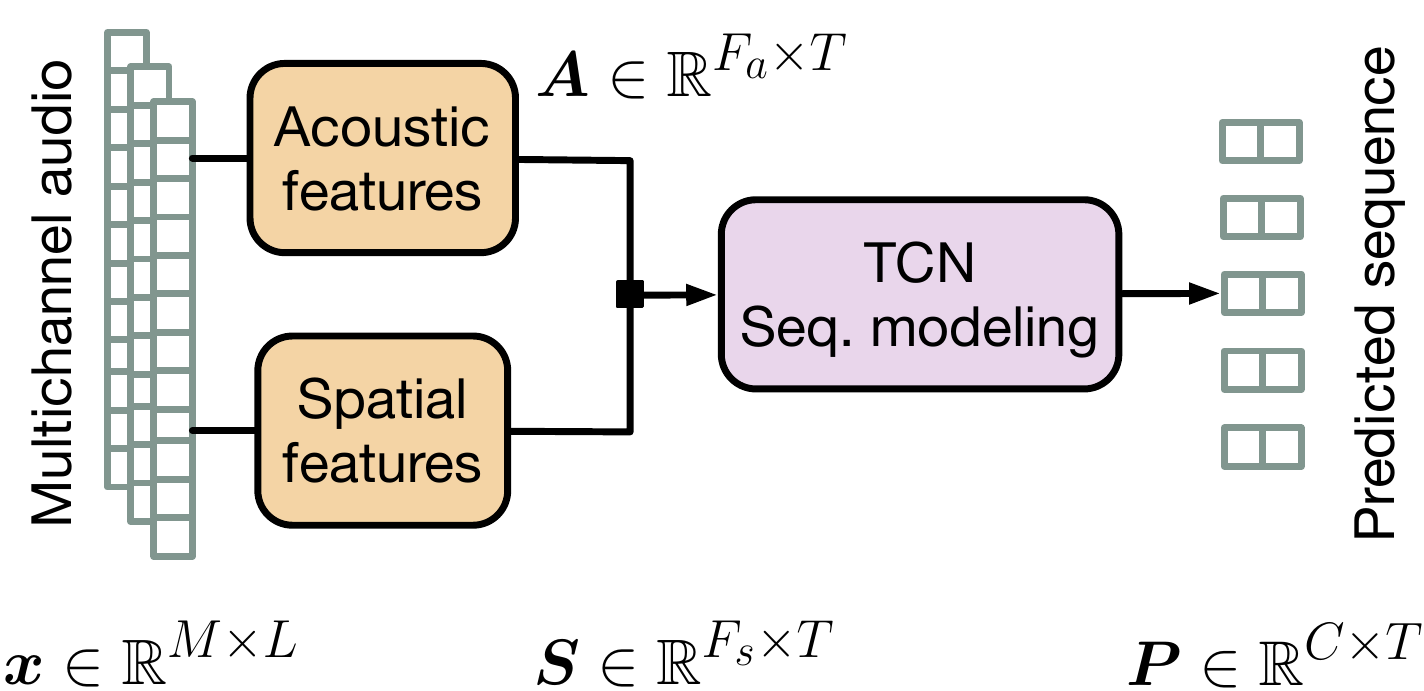}
    \caption{Segmentation architecture used for VAD, OSD, and SCD. Acoustic and spatial features are extracted from the multi-microphone signal. $\blacksquare$ denotes the concatenation operation. The model has two outputs ($C=2$) for classification and one output ($C=1$) for regression.}
    \label{fig:archi}
\end{figure}

The segmentation architecture -- figure~\ref{fig:archi} -- is composed of the following modules.
The acoustic feature module extracts a representation $\boldsymbol{A}\in\mathbb{R}^{F_a\times T}$ from the multi-microphone signal with $F_a$ being the acoustic feature size. 
The spatial feature module extracts a representation $\boldsymbol{S}\in\mathbb{R}^{F_s\times T}$ from the  same signal with $F_s$ being the spatial feature size. 
Both kinds of features are concatenated on the first dimension to produce a $F$-long feature vector.
The feature sequence is fed to the sequence modeling network which outputs the prediction $\boldsymbol{P}\in\mathbb{R}^{C\times T}$ with $C$ being the output size.

\vspace{-15pt}
\subsubsection{Acoustic features}
Acoustic features are extracted from the signal captured by the first channel of the microphone array.
Mel Frequency Cepstral Coefficients (MFCC) and Log-Mel spectrogram are used as acoustic features following~\cite{Cornell2022}.
Both types of features are extracted on a 25ms sliding window with a 10ms shift.
The mel spectrogram is obtained using 80 mel-scale filters before conversion to the log scale.
Time-frequency masking is applied as data augmentation during training.
This results in a vector of $F_a=80$ features.

20 MFCC are computed from a mel-spectrogram extracted using 40 triangular filters since it leads to the best performance.
The first MFCC coefficient (energy) is removed and both $\Delta$ and $\Delta\Delta$ are computed.
This results in a vector of $F_a=59$ features.

\subsubsection{Spatial features}

The proposed CH-DOA features are computed for each frequency bin of the STFT which results in a vector of $F_s=257$ features.
To ensure alignment with the acoustic features, spatial features are extracted on 25ms sliding windows with a 10ms shift.
Interaural Phase Difference (IPD) and cosine and sine IPD (CSIPD) are considered as baseline spatial features following~\cite{Cornell2022}.
In this work, we consider 4 microphone pairs with the microphones in opposition.
This results in a vector of $F_s=1028$ and $F_s=2056$ features respectively.

\subsubsection{Sequence modeling}
Sequence modeling is performed using the Temporal Convolutional Network\footnote{\url{https://github.com/popcornell/OSDC}}.
It consists of 1-d convolutional layers with exponentially increasing dilatation to learn a large temporal context.
Our architecture is composed of 3 stacked TCN blocks of 5 convolutional layers. 
A residual connection is added after each TCN block.
Before feeding the sequence of feature vectors to the TCN blocks, a 1-d bottleneck convolution compresses the feature sequence from $F$-dimensional vectors to 128 dimensions. 
Prediction is obtained by adding a 1-d convolution followed by a softmax activation function for the classification tasks or a linear activation in the case of regression (SCD).

\subsection{Training and evaluation procedures}

VAD, OSD, and SCD systems are trained on the AMI train subset. 
The classification models -- VAD and OSD -- are optimized using cross-entropy loss.
SCD models are optimized with the Mean Squared Error training objective.
Model weights are updated using the ADAM optimizer with a learning rate set to $lr=0.001$.
Each system is trained on 2s audio segments randomly sampled from the training set.
The batch size is set to 64.
Overlaps augmentation~\cite{bullock_overlap-aware_2020} is applied to 50\% of the training segments.
Models are trained on Nvidia RTX6000 GPU cards.

Models are evaluated on the AMI evaluation set.
Inference is performed on 2s segments with a 0.5s shift.
The VAD, OSD, and SCD detection thresholds are tuned on the development set.
VAD is evaluated in terms of False Alarm rate (FA) and missed detection (Miss).
OSD is evaluated using F1-score and average precision (AP)~\cite{cornell_detecting_2020}.
Finally, SCD is evaluated using purity ($P$) and coverage ($C$) as suggested in~\cite{bredin2017tristounet,yin2017speaker}.
The F1-score, i.e. the harmonic mean, of purity and coverage is also reported and is denoted as Segmentation Error (SER). 
We also report the 95\% confidence interval calculated on the file-level performance for each metric. 

%% file: 04_seg_results.tex
\section{Experimental study}
\label{sect4_seg_perf}

\begin{table*}[t!]
\centering
\caption{Performance of VAD, OSD, and SCD systems on the AMI meeting corpus with each type of features. The number of parameters is given for the VAD and OSD systems. The number of parameters of the regression model--SCD--is slightly lower. Bold values indicate the best-performing model for each type of acoustic feature. MFCC and Log-Mel correspond to single-channel models.}
\begin{tabular}{@{}lcccccccc@{}}
\toprule
 & & \multicolumn{2}{c}{VAD} &  \multicolumn{2}{c}{OSD} & \multicolumn{3}{c}{SCD}\\
\cmidrule(lr){3-4}
\cmidrule(lr){5-6}
\cmidrule(lr){7-9}
 & \# param.      & FA$_{\%\downarrow}$          & Miss$_{\%\downarrow}$ & F1-score$_{\%\uparrow}$ & AP$_{\%\uparrow}$ & $P_{\%\uparrow}$          & $C_{\%\uparrow}$ & $SER_{\%\uparrow}$\\
\midrule
MFCC & 0.26M & \textbf{3.5}$\pm$0.5 & \textbf{3.0}$\pm$1.3 & 64.5$\pm$5.6 & 65.1$\pm$7.2 & 82.2$\pm$2.2 & 79.2$\pm$2.6 & 80.7$\pm$0.9 \\
\quad + IPD & 0.33M & 4.2$\pm$0.8 & \textbf{3.0}$\pm$1.2 & 60.7$\pm$5.5 & 62.5$\pm$7.5 & 74.8$\pm$2.8 & 76.4$\pm$2.9 & 75.6$\pm$0.7\\
\quad + CSIPD & 0.40M & \textbf{3.0}$\pm$0.3 & 4.1$\pm$1.6 & \textbf{71.7}$\pm$4.5 & \textbf{75.9}$\pm$5.5 & 81.9$\pm$1.3 &\textbf{85.9}$\pm$2.7 & \textbf{83.9}$\pm$1.2 \\
\quad + CH-DOA & 0.28M & \textbf{3.4}$\pm$0.4 & \textbf{3.1}$\pm$1.4 & \textbf{69.3}$\pm$4.6 & \textbf{73.0}$\pm$5.8 & \textbf{84.6}$\pm$1.6 & \textbf{84.3}$\pm$3.4 & \textbf{84.4}$\pm$1.4 \\ \midrule
Log-Mel & 0.27M & \textbf{3.2}$\pm$0.4 & \textbf{3.5}$\pm$1.4 & 66.1$\pm$5.9 & 68.9$\pm$7.9 & 83.9$\pm$1.7 & 80.4$\pm$2.3 & 82.1$\pm$1.2 \\
\quad + IPD &  0.33M  & \textbf{3.0}$\pm$0.4 & 4.3$\pm$1.3 & 65.2$\pm$5.9  & 65.9$\pm$7.2 & 79.5$\pm$3.0 & 76.8$\pm$4.2 & 78.1$\pm$1.3\\
\quad + CSIPD & 0.40M & 3.7$\pm$0.4 & \textbf{3.1}$\pm$1.3 & \textbf{73.4}$\pm$5.3 & \textbf{75,6}$\pm$6.1 & 85.5$\pm$1.8 & \textbf{83.9}$\pm$3.9 & \textbf{84.7}$\pm$1.6\\
\quad + CH-DOA & 0.28M & \textbf{3.2}$\pm$0.4 & \textbf{3.3}$\pm$1.3 & 67.3$\pm$5.5 & 68.3$\pm$7.1 &\textbf{87.2}$\pm$1.4 & \textbf{82.5}$\pm$3.4 & \textbf{84.8}$\pm$1.5 \\
\bottomrule
\end{tabular}
\label{tab:all_res}
\end{table*}

This section presents the experimental results obtained with CH-DOA features on VAD, OSD, and SCD. 
The performance on each task is presented in table~\ref{tab:all_res}.

\subsection{VAD performance}

Considering only MFCC shows robust VAD performance with 3.5\% Miss and 3.0\% FA.
Adding IPD or CSIPD slightly degrades the performance as shown by the imbalanced Miss and FA.
The proposed CH-DOA shows similar performance as the MFCC with 3.4\% FA and 3.1\% Miss.
VAD performance is similar when considering Log-Mel features.
Using Log-Mel features offers 3.2\% FA and 3.5\% Miss.
Again, IPD and CSIPD fail at improving the VAD performance.
CH-DOA features show a similar performance as the Log-Mel with 3.2\% FA and 3.3\% Miss but without improvement.
Since the performance is the same between MFCC and CH-DOA, it seems the model does not use spatial information for VAD. 
Other information fusion schemes could be investigated instead of feature concatenation.
\vspace{-5pt}
\subsection{OSD performance}
Results show that adding IPD features (62.5\% AP) degrades OSD with regard to MFCC features (65.1\%).
This degradation can be seen in both F1-score and AP metrics.
Adding CSIPD features (75.9\% AP) significantly improves OSD performance with a +10.8\% absolute AP gain compared to MFCC.
Then, the proposed CH-DOA feature (73.0\% AP) reaches a similar performance as the CSIPD on both F1-score and AP with a little degradation.
This model, however, has only 0.28M parameters to optimize against 0.40M for the CSIPD one.

Models trained with Log-Mel features behave similarly to the MFCC with a global performance gain, except with CH-DOA features.
IPD features (65.9\% AP) degrade the OSD performance with regard to Log-Mel (68.9\% AP).
Again, CSIPD (75.6\% AP) offers the best OSD performance with a +6.7\% AP and a +7.3\% F1-score    absolute improvements with respect to Log-Mel.
In this configuration, CH-DOA features offer mitigated OSD performance (68.3\% AP) without improving nor degrading the detection.

\subsection{SCD performance}
SCD models behave similarly to OSD models. 
When MFCC features are considered, IPD degrades the SCD performance (75.6\% SER) with respect to MFCC (80.7\% SER).
Considering CSIPD features (83.9\% SER) significantly improves SCD performance with a +3.2\% absolute SER gain.
CH-DOA (84.4\% SER) reaches a similar performance, reaching a +3.7\% absolute SER gain.
This system also shows the best balance between purity and coverage.

Considering Log-Mel features with IPD degrades the detection by an absolute -4.0\% SER while both CSIPD and CH-DOA improve SER by +2.6\% and +2.7\% respectively.
The proposed CH-DOA features offer similar performance as CSIPD while reducing the number of trainable parameters.
Moreover, this model is not constrained to the array configuration used in the training data as shown in the following section.

\subsection{Robustness to the number of microphones}

CH-DOA is based on zero- and first-order circular harmonics. 
Hence, the feature extraction is not supposed to rely on the number of available microphones in the UCA.
This sub-section evaluates the two best-performing MFCC-based OSD and SCD models by desactivating 4 channels in the evaluation data.
Performance on OSD and SCD is presented in table~\ref{tab:4channels_only}.
Results on the OSD task show that CH-DOA features are more robust to a mismatch in the microphone number, reaching a 71.4\% AP.
This system remains better than MFCC (65.1\% AP) with an absolute +6.3\% AP improvement.
CSIPD features are less robust to array mismatch with a 51.5\% AP. 

On SCD, the CH-DOA model shows the best performance on both purity (84.1\%) and coverage (83.2\%) while still improving single-channel MFCC (82.2\%/79.2\%).
CSIPD features degrade the performance with $M=4$, mostly on coverage (75.8\%).
\vspace{-5pt}
\begin{table}[hb]
    \centering
    \caption{OSD and SCD performance on AMI array 1 evaluation data with $M=4$ deactivated channels.}
    \begin{adjustbox}{max width=\linewidth}
    \begin{tabular}{lccccc}
    \toprule
              & \multicolumn{2}{c}{OSD} & \multicolumn{2}{c}{SCD}\\
    \cmidrule(lr){2-3}
    \cmidrule(lr){4-5}
    $M=4$        & F1-score$_{\%\uparrow}$ & AP$_{\%\uparrow}$ & $P_{\%\uparrow}$ & $C_{\%\uparrow}$ \\ \midrule
         MFCC  & 64.5$\pm$5.6 & 65.1$\pm$7.2 & 82.2$\pm$2.2 & 79.2$\pm$2.6\\
         + CSIPD & 55.4$\pm6.7$ & 51.5$\pm$7.9& 81.1$\pm$1.7 & 75.8$\pm$ 3.0\\ 
         + CH-DOA & \textbf{69,6}$\pm$5.3 & \textbf{71.4}$\pm$6.3 & \textbf{84.1}$\pm$1.7 & \textbf{83.2}$\pm$3.4\\ \bottomrule
    \end{tabular}
    \end{adjustbox}
    \label{tab:4channels_only}
\end{table}
\vspace{-5pt}

%% file: 06_ccl.tex
\section{Conclusions}
\label{sect_ccl}

This paper introduces a new set of spatial features based on direction-of-arrival (DOA) estimation in the circular harmonics (CH) domain.
CH-DOA is investigated on three automatic speech segmentation tasks: Voice Activity Detection (VAD), Overlapped Speech Detection (OSD), and Speaker Change Detection (SCD).
The proposed CH-DOA is compared with state-of-the-art spatial features and combined with commonly used acoustic features.
Although limited to circular arrays, CH-DOA shows better segmentation performance than single-channel acoustic features, particularly on OSD and SCD.
Furthermore, we demonstrate that adding spatial features significantly improves SCD and reach the best performance with CH-DOA (84.8\% SER).
Finally, CH-DOA shows encouraging robustness to array mismatch by still improving SCD and OSD under these conditions.

The use of information fusion schemes (e.g. cross attention) will be investigated in future work since spatial information seems less exploited on the VAD.
The segmentation models remain to be evaluated in a full diarization pipeline.

%% file: main.bbl
\begin{thebibliography}{10}
\providecommand{\url}[1]{#1}
\csname url@samestyle\endcsname
\providecommand{\newblock}{\relax}
\providecommand{\bibinfo}[2]{#2}
\providecommand{\BIBentrySTDinterwordspacing}{\spaceskip=0pt\relax}
\providecommand{\BIBentryALTinterwordstretchfactor}{4}
\providecommand{\BIBentryALTinterwordspacing}{\spaceskip=\fontdimen2\font plus
\BIBentryALTinterwordstretchfactor\fontdimen3\font minus
  \fontdimen4\font\relax}
\providecommand{\BIBforeignlanguage}[2]{{%
\expandafter\ifx\csname l@#1\endcsname\relax
\typeout{** WARNING: IEEEtran.bst: No hyphenation pattern has been}%
\typeout{** loaded for the language `#1'. Using the pattern for}%
\typeout{** the default language instead.}%
\else
\language=\csname l@#1\endcsname
\fi
#2}}
\providecommand{\BIBdecl}{\relax}
\BIBdecl

\bibitem{anguera2012speaker}
X.~Anguera, S.~Bozonnet, N.~Evans, C.~Fredouille, G.~Friedland, and O.~Vinyals,
  ``Speaker diarization: A review of recent research,'' \emph{IEEE Transactions
  on audio, speech, and language processing}, vol.~20, no.~2, pp. 356--370,
  2012.

\bibitem{park2022review}
T.~J. Park, N.~Kanda, D.~Dimitriadis, K.~J. Han, S.~Watanabe, and S.~Narayanan,
  ``A review of speaker diarization: Recent advances with deep learning,''
  \emph{Computer Speech \& Language}, vol.~72, p. 101317, 2022.

\bibitem{bredin_pyannoteaudio_2020}
H.~Bredin, R.~Yin, J.~M. Coria, G.~Gelly, P.~Korshunov, M.~Lavechin, D.~Fustes,
  H.~Titeux, W.~Bouaziz, and M.-P. Gill,
  ``\BIBforeignlanguage{en}{Pyannote.{Audio}: {Neural} {Building} {Blocks} for
  {Speaker} {Diarization}},'' in \emph{\BIBforeignlanguage{en}{ICASSP}}, 2020,
  pp. 7124--7128.

\bibitem{landini2022bayesian}
F.~Landini, J.~Profant, M.~Diez, and L.~Burget, ``Bayesian hmm clustering of
  x-vector sequences (vbx) in speaker diarization: theory, implementation and
  analysis on standard tasks,'' \emph{Computer Speech \& Language}, vol.~71, p.
  101254, 2022.

\bibitem{bredin21_interspeech}
H.~Bredin and A.~Laurent, ``{End-To-End Speaker Segmentation for Overlap-Aware
  Resegmentation},'' in \emph{Proc. Interspeech 2021}, 2021, pp. 3111--3115.

\bibitem{garcia2019speaker}
P.~Garc{\'\i}a, J.~Villalba, H.~Bredin, J.~Du, D.~Castan, A.~Cristia,
  L.~Bullock, L.~Guo, K.~Okabe, P.~S. Nidadavolu \emph{et~al.}, ``Speaker
  detection in the wild: Lessons learned from jsalt 2019,'' \emph{arXiv
  preprint arXiv:1912.00938}, 2019.

\bibitem{sohn1999statistical}
J.~Sohn, N.~S. Kim, and W.~Sung, ``A statistical model-based voice activity
  detection,'' \emph{IEEE signal processing letters}, vol.~6, no.~1, pp. 1--3,
  1999.

\bibitem{ng2012developing}
T.~Ng, B.~Zhang, L.~Nguyen, S.~Matsoukas, X.~Zhou, N.~Mesgarani, K.~Vesel{\`y},
  and P.~Mat{\v{e}}jka, ``Developing a speech activity detection system for the
  darpa rats program,'' in \emph{Thirteenth annual conference of the
  international speech communication association}, 2012.

\bibitem{boakye_improved_2011}
K.~Boakye, O.~Vinyals, and G.~Friedland, ``\BIBforeignlanguage{en}{Improved
  overlapped speech handling for speaker diarization},'' in
  \emph{\BIBforeignlanguage{en}{Interspeech}}, 2011, pp. 941--944.

\bibitem{charlet_impact_2013}
D.~Charlet, C.~Barras, and J.-S. Lienard, ``\BIBforeignlanguage{en}{Impact of
  overlapping speech detection on speaker diarization for broadcast news and
  debates},'' in \emph{\BIBforeignlanguage{en}{ICASSP}}, 2013, pp. 7707--7711.

\bibitem{yella2014overlapping}
S.~H. Yella and H.~Bourlard, ``Overlapping speech detection using long-term
  conversational features for speaker diarization in meeting room
  conversations,'' \emph{IEEE/ACM Transactions on Audio, Speech, and Language
  Processing}, vol.~22, no.~12, pp. 1688--1700, 2014.

\bibitem{siegler1997automatic}
M.~A. Siegler, U.~Jain, B.~Raj, and R.~M. Stern, ``Automatic segmentation,
  classification and clustering of broadcast news audio,'' in \emph{Proc. DARPA
  speech recognition workshop}, vol. 1997, 1997.

\bibitem{liu1999fast}
D.~Liu and F.~Kubala, ``Fast speaker change detection for broadcast news
  transcription and indexing,'' in \emph{Sixth European Conference on Speech
  Communication and Technology}, 1999.

\bibitem{ryant2013speech}
N.~Ryant, M.~Liberman, and J.~Yuan, ``Speech activity detection on youtube
  using deep neural networks.'' in \emph{INTERSPEECH}.\hskip 1em plus 0.5em
  minus 0.4em\relax Lyon, France, 2013, pp. 728--731.

\bibitem{lavechin2019end}
M.~Lavechin, M.-P. Gill, R.~Bousbib, H.~Bredin, and L.~P. Garcia-Perera,
  ``{End-to-End Domain-Adversarial Voice Activity Detection},'' in \emph{Proc.
  Interspeech 2020}, 2020, pp. 3685--3689.

\bibitem{bullock_overlap-aware_2020}
L.~Bullock, H.~Bredin, and L.~P. Garcia-Perera,
  ``\BIBforeignlanguage{en}{Overlap-{Aware} {Diarization}: {Resegmentation}
  {Using} {Neural} {End}-to-{End} {Overlapped} {Speech} {Detection}},'' in
  \emph{\BIBforeignlanguage{en}{ICASSP}}, 2020, pp. 7114--7118.

\bibitem{Cornell2022}
\BIBentryALTinterwordspacing
S.~Cornell, M.~Omologo, S.~Squartini, and E.~Vincent, ``Overlapped speech
  detection and speaker counting using distant microphone arrays,''
  \emph{Computer Speech {\&} Language}, vol.~72, p. 101306, 2022. [Online].
  Available: \url{https://doi.org/10.1016/j.csl.2021.101306}
\BIBentrySTDinterwordspacing

\bibitem{bredin2017tristounet}
H.~Bredin, ``Tristounet: triplet loss for speaker turn embedding,'' in
  \emph{2017 IEEE international conference on acoustics, speech and signal
  processing (ICASSP)}.\hskip 1em plus 0.5em minus 0.4em\relax IEEE, 2017, pp.
  5430--5434.

\bibitem{yin2017speaker}
R.~Yin, H.~Bredin, and C.~Barras, ``Speaker change detection in broadcast tv
  using bidirectional long short-term memory networks,'' in \emph{Interspeech
  2017}.\hskip 1em plus 0.5em minus 0.4em\relax ISCA, 2017.

\bibitem{hruz2017convolutional}
M.~Hr{\'u}z and Z.~Zaj{\'\i}c, ``Convolutional neural network for speaker
  change detection in telephone speaker diarization system,'' in \emph{2017
  IEEE International Conference on Acoustics, Speech and Signal Processing
  (ICASSP)}.\hskip 1em plus 0.5em minus 0.4em\relax IEEE, 2017, pp. 4945--4949.

\bibitem{hu2015speaker}
M.~Hu, D.~Sharma, S.~Doclo, M.~Brookes, and P.~A. Naylor, ``Speaker change
  detection and speaker diarization using spatial information,'' in \emph{2015
  IEEE International Conference on Acoustics, Speech and Signal Processing
  (ICASSP)}.\hskip 1em plus 0.5em minus 0.4em\relax IEEE, 2015, pp. 5743--5747.

\bibitem{cornell_detecting_2020}
S.~Cornell, M.~Omologo, S.~Squartini, and E.~Vincent,
  ``\BIBforeignlanguage{en}{Detecting and {Counting} {Overlapping} {Speakers}
  in {Distant} {Speech} {Scenarios}},'' in
  \emph{\BIBforeignlanguage{en}{Interspeech}}, 2020, pp. 3107--3111.

\bibitem{mariotte22_interspeech}
T.~Mariotte, A.~Larcher, S.~Montrésor, and J.-H. Thomas, ``{Microphone Array
  Channel Combination Algorithms for Overlapped Speech Detection},'' in
  \emph{Proc. Interspeech 2022}, 2022, pp. 4636--4640.

\bibitem{pak2019sound}
J.~Pak and J.~W. Shin, ``Sound localization based on phase difference
  enhancement using deep neural networks,'' \emph{IEEE/ACM Transactions on
  Audio, Speech, and Language Processing}, vol.~27, no.~8, pp. 1335--1345,
  2019.

\bibitem{vincent2018audio}
E.~Vincent, T.~Virtanen, and S.~Gannot, \emph{Audio source separation and
  speech enhancement}.\hskip 1em plus 0.5em minus 0.4em\relax John Wiley \&
  Sons, 2018.

\bibitem{gu2020enhancing}
R.~Gu, S.-X. Zhang, L.~Chen, Y.~Xu, M.~Yu, D.~Su, Y.~Zou, and D.~Yu,
  ``Enhancing end-to-end multi-channel speech separation via spatial feature
  learning,'' in \emph{ICASSP 2020-2020 IEEE International Conference on
  Acoustics, Speech and Signal Processing (ICASSP)}.\hskip 1em plus 0.5em minus
  0.4em\relax IEEE, 2020, pp. 7319--7323.

\bibitem{grumiaux2021high}
P.-A. Grumiaux, S.~Kiti{\'c}, L.~Girin, and A.~Gu{\'e}rin, ``High-resolution
  speaker counting in reverberant rooms using crnn with ambisonics features,''
  in \emph{2020 28th European Signal Processing Conference (EUSIPCO)}.\hskip
  1em plus 0.5em minus 0.4em\relax IEEE, 2021, pp. 71--75.

\bibitem{songgong_robust_2021}
\BIBentryALTinterwordspacing
K.~SongGong and H.~Chen, ``\BIBforeignlanguage{en}{Robust {Indoor} {Speaker}
  {Localization} in the {Circular} {Harmonic} {Domain}},''
  \emph{\BIBforeignlanguage{en}{IEEE Transactions on Industrial Electronics}},
  vol.~68, no.~4, pp. 3413--3422, Apr. 2021. [Online]. Available:
  \url{https://ieeexplore.ieee.org/document/9037178/}
\BIBentrySTDinterwordspacing

\bibitem{carletta2006ami}
J.~Carletta, S.~Ashby, S.~Bourban, M.~Flynn, M.~Guillemot, T.~Hain, J.~Kadlec,
  V.~Karaiskos, W.~Kraaij, M.~Kronenthal \emph{et~al.}, ``The ami meeting
  corpus: A pre-announcement,'' in \emph{Machine Learning for Multimodal
  Interaction: Second International Workshop, MLMI 2005, Edinburgh, UK, July
  11-13, 2005, Revised Selected Papers 2}.\hskip 1em plus 0.5em minus
  0.4em\relax Springer, 2006, pp. 28--39.

\bibitem{fu21b_interspeech}
Y.~Fu, L.~Cheng, S.~Lv, Y.~Jv, Y.~Kong, Z.~Chen, Y.~Hu, L.~Xie, J.~Wu, H.~Bu,
  X.~Xu, J.~Du, and J.~Chen, ``{AISHELL-4: An Open Source Dataset for Speech
  Enhancement, Separation, Recognition and Speaker Diarization in Conference
  Scenario},'' in \emph{Proc. Interspeech 2021}, 2021, pp. 3665--3669.

\bibitem{he_deep_2018}
W.~He, P.~Motlicek, and J.-M. Odobez, ``Deep neural networks for multiple
  speaker detection and localization,'' in \emph{2018 IEEE International
  Conference on Robotics and Automation (ICRA)}.\hskip 1em plus 0.5em minus
  0.4em\relax IEEE, 2018, pp. 74--79.

\bibitem{williams2000fourier}
E.~G. Williams and J.~A. Mann~III, ``Fourier acoustics: sound radiation and
  nearfield acoustical holography,'' 2000.

\bibitem{torres_room_2016}
A.~M. Torres, J.~Mateo, and M.~Cobos, ``\BIBforeignlanguage{en}{Room
  {Acoustics} {Analysis} {Using} {Circular} {Arrays}: {A} {Comparison}
  {Between} {Plane}-{Wave} {Decomposition} and {Modal} {Beamforming}
  {Approaches}},'' \emph{\BIBforeignlanguage{en}{Circuits, Systems, and Signal
  Processing}}, vol.~35, no.~5, pp. 1625--1642, May 2016.

\end{thebibliography}
